\documentclass[11pt]{article}

\catcode`\@=11
\@addtoreset{equation}{section}

\evensidemargin \oddsidemargin

\usepackage{mathrsfs,amsbsy,amssymb,latexsym,amsfonts,amsmath,
amssymb}

\usepackage{cite}

\usepackage{color}
  \usepackage{ascmac}
  
  \usepackage{bm}

\usepackage[dvipdfmx]{hyperref}

\hypersetup
{setpagesize=false,
 bookmarksnumbered=true,
 bookmarksopen=true,
 colorlinks=true,
 linkcolor=black,
 citecolor=black,}

\newcommand\pa{\partial}

\newcommand\nn{\nonumber}

\newcommand\phis{{\phi^*}}

\newcommand\tr{{\rm tr}\,}

\newcommand\diag{{\rm diag}}

\newcommand\bref[1]{(\ref{#1})}

\begin{document}

\hfill 
\vspace{3cm}

\begin{center}
{\LARGE \bf
A Comment on the Higher-Spin Gauge Models
}
\vspace{40mm}\\
Ryota Fujii
\,and~
Makoto Sakaguchi

\vspace{15mm}

Department of Physics, Ibaraki University, Mito 310-8512, Japan
\end{center}

\vspace{25mm}

\begin{abstract}
We examine the higher-spin gauge models which are free on-shell
and whose cubic vertex is BRST-exact.
We show that these are free off-shell as well as on-shell.
The key equation for this relates the derivative of the total deformed action $S(g)$ with respect to the deformation parameter $g$
to an $S(g)$-exact term.

\end{abstract}

\thispagestyle{empty}
\setcounter{page}{0}

\newpage

\setcounter{footnote}{0}

Higher-spin gauge theories are studied from various viewpoints.
In the realm of  string theory,
they are expected to reveal characteristic features in the high-energy limit\cite{Gross} (see also \cite{Sagnotti tensionless limit}).
Free higher-spin gauge theories are now well understood,
while interacting theories have not been understood well
even for massless higher-spin gauge fields in flat spacetime.
There are several obstacles to construct intractions, one of which is the no-go theorem\cite{no-go}\footnote{
See \cite{KU81},
for an analysis in the field theory framework.}.
To avoid this,
the spin value and the number of the derivatives contained in vertices should be restricted so that higher-spin gauge fields are not included in the asymptotic states.

 \medskip

By applying the BRST deformation scheme\cite{BH9304}
in the BRST-antifield formalism,
higher-spin gauge models of massless totally-symmetric tensors
are constructed in \cite{SS2011}\footnote{
See \cite{BDGH0007, BRST cohomology,HLGR1206,HLGR1310, Rahman1905} for other attempts by using BRST cohomological metod.
}.
Subsequently,
these are generalized to include massless totally-symmetric tensor-spinors in \cite{FKSS2024}.
As an example,
we consider
the higher-spin gauge model of two bosons,
a totally-symmetric rank-$n_1$ tensor $\phi^{\mu_1\cdots\mu_{n_1}}_1$ and a rank-$n$ tensor $\phi^{\mu_1\cdots\mu_n}$.
We also introduce
Grassmann-odd ghost fields, 
a rank-$(n_1-1)$ tensor $c_1^{\mu_2\cdots\mu_{n_1}}$ and a rank-$(n-1)$ tensor $c^{\mu_2\cdots\mu_n}$,
and antifields  $\{\phi^*_1, \phi^*, c_1^*,  c^*\}$
which have the same algebraic symmetries but opposite Grassmann parity.
The action is
given as
\begin{align}
\label{def. action}S(g)\ =\ S^0+\sum^\infty_{n=1}g^nS^n,
\end{align}
where  $g$ is the deformation parameter.
$S^0$ represents the free action,
 while 
$S^n$ ($n\geq 1$) does
the ($n+2$)-point vertex
composed of $n+2$ (anti)fields,
given by\footnote{
Hereafter,
to simplify,
we frequently omit totally-symmetric indices but this may not cause confusion.}
\begin{align}
\label{free action}S^0\ =&\int d^{D}x\ \text{tr}\Bigg[\frac{1}{2}\phi_1^TG(\phi_{1})
+\phi^*_1{}^T\partial c_1
+\frac{1}{2}\phi^TG(\phi)+\phi^*{}^T\partial c\Bigg],
\\
S^n\ =& \int d^{D}x\ \text{tr}\Bigg[
\frac{1}{2}\phi^T_1\Phi^n[G(\phi_1)]
+(\tilde{\partial}c\phi^*_1)^T\Phi^{n-1}[G(\phi_1)]
+\frac{1}{2}(\tilde{\partial}c\phi^*_1)^T\Phi^{n-2}[\tilde{G}(\tilde{\partial}c\phi^*_1)]
\Bigg],
\label{eqn:Sn}
\end{align}
where $\Phi^0[A]=A,\ \Phi[A]=\tilde{G}(\phi A),\ \Phi^2[A]=\tilde{G}(\phi\tilde{G}(\phi A))$,
and so on.
For an arbitrary totally-symmetric rank-$n$ tensor $A$, we have defined $\tilde{G}(A)$ 
as $\tilde{G}(A)\equiv G(A)+\frac{1}{2}\eta\partial\partial A''$ 
and $G(A)$ as $G(A)\equiv F(A)-\frac{1}{2}\eta F'(A)$.
$F$ is the Fronsdal tensor \cite{Fronsdal78}
 defined by\footnote{
Our convention for symmetrization of indices is
\begin{align}
\pa_{(\mu_1}\cdots\pa_{\mu_r}A_{\mu_{r+1}\cdots\mu_n)\nu_1\cdots\nu_m}
=\frac{1}{r! (n-r)!}\sum_{\{\mu_1,\ldots,\mu_n\}}
\pa_{\mu_1}\cdots\pa_{\mu_r}A_{\mu_{r+1}\cdots\mu_n\nu_1\cdots\nu_m}
\nn,
\end{align}
where $\{\mu_1,\ldots,\mu_n\}$ indicates that 
the sum is taken over permutations of $\mu_1,\ldots,\mu_n$.}
\begin{align}
F_{\mu_1\cdots\mu_n}(A)=&\square A_{\mu_1\cdots\mu_n}
-\pa_{(\mu_1}\pa\cdot A_{\mu_2\cdots\mu_n)}
+\pa_{(\mu_1}\pa_{\mu_2}A'_{\mu_3\cdots\mu_n)},
\label{Fronsdal bos flat}
\end{align}
where $\square=\eta^{\mu\nu}\pa_\mu\pa_\nu$
and $\eta^{\mu\nu}=\diag(-1,+1,\ldots,+1)$.
A prime on a tensor represents the trace, 
namely $A'_{\mu_3\cdots \mu_n}=A_{\mu_3\cdots \mu_n}{}^\rho{}_\rho$\,.
The divergence of $A$
is expressed as
$\pa\cdot A_{\mu_2\cdots\mu_n}=\pa^{\mu_1}A_{\mu_1\mu_2\cdots\mu_n}$\,.
We require that the ghost $c$ is traceless for $n\geq3$,
so that $F(\phi)$ is invariant under the BRST transformation $\delta_B\phi=\partial c$.
For notational simplicity,
we used a matrix notation in \bref{free action} and \bref{eqn:Sn}.
For example,
the integrand of the second term in \bref{eqn:Sn} at $n=1$ means
\begin{align}
\tilde{\partial}_{(\mu_1}c_{\mu_2\cdots\mu_q)\rho_1\cdots\rho_q}
\phis^{\rho_1\cdots\rho_q\nu_1\cdots\nu_p}_1
G_{\nu_1\cdots\nu_p}{}^{\mu_1\cdots\mu_q}(\phi_1).
\end{align}
Here $p+q=n_1$ and $2q=n$.
The $\phi_1$ and $\phis_1$ denote $d(p)\times d(q)$ matrices.

In \cite{FKSS2024}, this model is shown to be free on-shell.
Furthermore,
it is shown that the cubic vertex is BRST-exact
\begin{align}
S^1\ =\delta_BR^1\ =\ (R^1,S^0),\qquad
R^1=-\frac{1}{2}\int\ d^Dx\ \tr\Big[
\phi_1^T\tilde{G}(\phi\phi^*_1)
+(\tilde\partial c\phi_1^*)^T\phi^*_1
\Big],
\label{eqn:R1}
\end{align}
where the antibracket
$(X,Y)$ for two functionals, $X[\Phi^A,\Phi^*_A]$ and $Y[\Phi^A,\Phi^*_A]$,
is defined by
\begin{align}
(X,Y)\equiv
X\frac{\overleftarrow{\delta}}{\delta\Phi^A}\frac{\overrightarrow{\delta}}{\delta\Phi^*_A}Y
+X\frac{\overleftarrow{\delta}}{\delta\Phi^*_A}\frac{\overrightarrow{\delta}}{\delta\Phi^A}Y,
\end{align}
where $\Phi^*_A$ are the antifields corresponding to $\Phi^A=\{\phi_1, \phi, c_1, c\}$.
However, whether the deformed total action $S(g)$ is free off-shell or not was left as an interesting future problem.
In this letter,
we tackle this issue and show that this higher-spin gauge model
reduces to free off-shell as well.

\subsection*{Reduction of  $S(g)$ to $S^0$ }

For our purpose, we show that there exists
$\mathfrak{R}(g)$ satisfying the differential equation
\begin{align}
\label{diff. eq.}\frac{d}{dg}S(g)\ =\ \Big(S(g),\mathfrak{R}(g)\Big),
\end{align}
which means that  the derivative of $S(g)$ with respect to $g$ is $S$-exact.
This is  the key relation to relate $S(g)$ to $S^0$.
Expanding $\mathfrak{R}(g)$ in terms of $g$ as
\begin{align}
\mathfrak{R}(g)\ =\ \sum^\infty_{n=1}g^{n-1}\mathfrak{R}^n,
\end{align}
the differential equation (\ref{diff. eq.}) reduces to
\begin{align}
\label{SST}nS^n\ =\ \Big(S^0,\mathfrak{R}^n\Big)+\sum^{n-1}_{k=1}\Big(S^{k},\mathfrak{R}^{n-k}\Big),
\end{align}
at the order of $g^{n-1}$ ($n\geq 1$),
or equivalently 
\begin{align}
\label{SST1}S^1\ =&\ \Big(S^0,\mathfrak{R}^1\Big),\\
\label{SST2}2S^2\ =&\ \Big(S^0,\mathfrak{R}^2\Big)+\Big(S^1,\mathfrak{R}^1\Big),\\
\label{SST3}3S^3\ =&\ \Big(S^0,\mathfrak{R}^3\Big)+\Big(S^1,\mathfrak{R}^2\Big)+\Big(S^2,\mathfrak{R}^1\Big),\\
\vdots&.\nonumber
\end{align}
As was seen in \bref{eqn:R1},
\bref{SST1} is satisfied 
by
\begin{align}
\mathfrak{R}^1\ 
=\ -R^1\ 
=\ \frac{1}{2}\int\ d^Dx\ \tr\Big[
\phi_1^T\tilde{G}(\phi\phi^*_1)
+(\tilde\partial c\phi_1^*)^T\phi^*_1
\Big].
\end{align}
Now, we show that the solution of  (\ref{SST}) is given by
\begin{align}
\mathfrak{R}^n\ =\ \frac{1}{2}\int d^{D}x\ {\rm{tr}}\Bigg[
\phi^T_1\Phi^{n}[\phi^*_1]
+(\tilde{\partial}c\phi^*_1)^T\Phi^{n-1}[\phi^*_1]
\Bigg].\label{solution}
\end{align}
Noting that
\begin{align}
\Big(S^k,\mathfrak{R}^{n-k}\Big)
\ =&\  \frac{1}{2}\int d^{D}x\ {\rm{tr}}\Bigg[
\Phi^k[G(\phi_1)]^T\phi\Phi^{n-k-1}[G(\phi_1)]
-\Phi^k[G(\phi_1)]^T(\tilde{\partial}c)^T\Phi^{n-k-1}[\phi^*_1]\nonumber\\
&\hspace{10mm}+\Phi^k[G(\phi_1)]^T\phi\Phi^{n-k-2}[\tilde{G}(\tilde{\partial}c\phi^*_1)]
-\Phi^{k-1}[G(\phi_1)]^T\tilde{\partial}c\Phi^{n-k}[\phi^*_1]\nonumber\\
&\hspace{10mm}+\Phi^{k-1}[\tilde{G}(\tilde{\partial}c\phi^*_1)]^T\phi\Phi^{n-k-1}[G(\phi_1)]
-\Phi^{k-1}[\tilde{G}(\tilde{\partial}c\phi^*_1)]^T(\tilde{\partial}c)^T\Phi^{n-k-1}[\phi^*_1]\nonumber\\
&\hspace{10mm}+\Phi^{k-1}[\tilde{G}(\tilde{\partial}c\phi^*_1)]^T\phi\Phi^{n-k-2}[G(\tilde{\partial}c\phi^*_1)]
-\Phi^{k-2}[\tilde{G}(\tilde{\partial}c\phi^*_1)]^T\tilde{\partial}c\Phi^{n-k}[\phi^*_1]
\Bigg],
\label{eqn:(Sk,Rn-k)}
\end{align}
we obtain
\begin{align}
\sum^{n-1}_{k=1}\Big(S^k,\mathfrak{R}^{n-k}\Big)
=&\ (n-1)S^n\nn
\\&
-\frac{1}{2}\int d^{D}x\ {\rm{tr}}\Bigg[
G(\phi_1)^T\tilde{\partial}c\Phi^{n-1}[\phi^*_1]
+\Phi^{n-1}[G(\phi_1)]^T(\tilde{\partial}c)^T\phi^*_1\nonumber\\
&\hspace{25mm}+\sum^{n-2}_{k=1}\Phi^k[G(\phi_1)]^T\partial c\Phi^{n-k-1}[\phi^*_1]
+\Phi^{n-2}[\tilde{G}(\tilde{\partial}c\phi^*_1)]^T(\tilde{\partial}c)^T\phi^*_1\nonumber\\
&\hspace{25mm}+\sum^{n-2}_{k=1}\Phi^{k-1}[\tilde{G}(\tilde{\partial}c\phi^*_1)]^T\partial c\Phi^{n-k-1}[\phi^*_1]\Bigg].
\end{align}
On the other hand, the antibraket between  $S^0$ and $\mathfrak{R}^n$ yields
\begin{align}
\Big(S^0,\mathfrak{R}^n\Big)
\ =&\ S^n
+\frac{1}{2}\int d^{D}x\ {\rm{tr}}\Bigg[
G(\phi_1)^T\tilde{\partial}c\Phi^{n-1}[\phi^*_1]
+\Phi^{n-1}[G(\phi_1)]^T(\tilde{\partial}c)^T\phi^*_1\nonumber\\
&\hspace{30mm}+\sum^{n-2}_{k=1}\Phi^k[G(\phi_1)]^T\partial c\Phi^{n-k-1}[\phi^*_1]
+\Phi^{n-2}[\tilde{G}(\tilde{\partial}c\phi^*_1)]^T(\tilde{\partial}c)^T\phi^*_1\nonumber\\
&\hspace{30mm}+\sum^{n-2}_{k=1}\Phi^{k-1}[\tilde{G}(\tilde{\partial}c\phi^*_1)]^T\partial c\Phi^{n-k-1}[\phi^*_1]\Bigg].
\label{eqn:(S0,Rn)}
\end{align}
Gathering \bref{eqn:(Sk,Rn-k)} and  \bref{eqn:(S0,Rn)} together,
we find that \bref{solution} solves \bref{SST}.

Now we are ready to show that $S(g)$ reduces to $S^0$.
By an infinitesimal change of $g$: $g\mapsto g+\delta g$,
the deformed action $S(g)$ becomes
\begin{align}\label{infinitesimal tr.}
S(g+\delta g)\ 
\simeq&\ S(g)+\delta g\frac{d}{dg}S(g).
\end{align}
Since the second term in the right hand side is $S$-exact thanks to
 the differential equation (\ref{diff. eq.}),
it can be eliminated by field redefinitions.
In fact,
the field redefinition $\Phi^A\mapsto \Phi^A+\delta gu^A$ and $\Phi^*_A\mapsto \Phi^*_A+\delta gv_A$ causes the change $S[\Phi^A,\Phi^*_A]\mapsto S[\Phi^A+gu^A,\Phi^*_A+gv_A]=S[\Phi^A,\Phi^*_A]+\delta g\Big(u^A\frac{\delta S}{\delta\Phi^A}+v_A\frac{\delta S}{\delta\Phi^*_A}\Big)$,
which can absorb $\frac{d}{dg}S(g)$ by choosing $u^A$ and $v_A$ appropriately.
As a result,
we obtain 
\begin{align}
S(g+\delta g)\ =\ S(g),
\end{align}
by field redefinitions.
This implies that the deformed action $S(g)$ is independent of $g$.
Therefore,
we may conclude that 
\begin{align}
S(g)\ =\ S(0)\ =\ S^0.
\end{align}

\bigskip

We have shown that the higher-spin gauge model (\ref{def. action}) 
with \bref{free action} and \bref{eqn:Sn} is  free off-shell as well as on-shell.
The key relation is \bref{diff. eq.}.
Other models in \cite{SS2011,FKSS2024} 
may satisfy the similar relation
and also be shown to be free off-shell by the similar argument.

Finally, we will comment on how $S^1$ becomes exact.
The cubic vertex $S^1$ in \cite{FKSS2024,SS2011} is given as $S^1=\alpha^1_0+\alpha^1_1+\alpha^1_2$ where the antighost number of $\alpha_i^1$ is $i$.
The master equation $(S(g),S(g))=0$
implies that 
 $\alpha^1_i$ must solve the relations
\begin{align}
\Gamma \alpha^1_2\ =\ 0,\qquad
\Delta \alpha^1_2+\Gamma \alpha^1_1\ =\ 0,\qquad
\Delta \alpha^1_1+\Gamma \alpha^1_0\ =\ 0.
\end{align}
The $\Delta$ and $\Gamma$ denote the BRST transformation that change the antighost number by $-1$ and $0$,
respectively.
On the other hand, $S^1$ should not be BRST-exact.
When $S^1$ is BRST-exact,
$\alpha^1_i$ may be expressed as
\begin{align}
\alpha^1_2\ =\ \Gamma r_2,\qquad
\alpha^1_1\ =\ \Delta r_2 +\Gamma r_1,\qquad
\alpha^1_0\ =\ \Delta r_1+\Gamma r_0.
\end{align}
In constructing higher-spin gauge models,  $\alpha_2^1=0$ was imposed,
but this does not mean $r_2=0$.
In fact, there may exist a non-trivial $r_2$ which satisfys $\Gamma r_1=0$
and $\Delta r_2\neq 0$.
The cubic vertex $S^1$ in \bref{eqn:R1} yields such an example.

\medskip

It is interesting to 
to consider the other cases,
including non-cubic vertex
$S^1$,
including massive tensor and tensor-spinor,
including mixed-symmetry (or, totally-antisymmetry) tensor and tensor-spinor,
and so on.
There may be a possibility to construct non-trivial interactions in these cases.

\section*{Acknowledgments}
The authors would like to thank Hiraki Kanehisa for useful comments.
This work was supported by JSPS KAKENHI Grant Number
JP21K03566
and
JST,
the establishment of university fellowships towards the creation of science technology innovation, 
Grant Number
JPMJFS2105.


\end{document}